\documentclass[aps,prd,floats,nofootinbib]{revtex4}
\usepackage{graphicx}

\begin{document}
\preprint{
\hfill
\begin{minipage}[t]{3in}
\begin{flushright}
\vspace{-0.2in}
FERMILAB--PUB--07--XXX--A
\end{flushright}
\end{minipage}
}
\bigskip

\title{The Intergalactic Propagation of Ultra-High Energy
Cosmic Ray Nuclei: an Analytic Approach\\}
\author{Dan Hooper$^{1}$, Subir Sarkar$^2$ and Andrew M.~Taylor$^3$}
\affiliation{
$^1$Theoretical Astrophysics Group, Fermilab, Batavia, IL  60510, USA\\
$^2$ Rudolf Peierls Centre for Theoretical Physics, University of Oxford, 
     1 Keble Road, Oxford OX1 3NP, UK\\
$^3$ Max-Planck-Institut f\"{u}r Kernphysik, Postfach 103980, D-69029,
Heidelberg, Germany}
\date{\today}

\bigskip

\begin{abstract}
  It is likely that ultra-high energy cosmic rays contain a
  significant component of heavy or intermediate mass nuclei. The
  propagation of ultra-high energy nuclei through cosmic radiation
  backgrounds is more complicated than that of protons and its study
  has required the use of Monte Carlo techniques. We present an
  analytic method for calculating the spectrum and the composition at
  Earth of ultra-high energy cosmic rays which start out as heavy
  nuclei from their extragalactic sources. The results obtained are in
  good agreement with those obtained using numerical methods.
\end{abstract}

\maketitle

\section{Introduction}

Recent observations by the High Resolution Fly's Eye~\cite{HiResSpec}
and the Pierre Auger Observatory (PAO)~\cite{AugerSpec} have
established that at energies above $\sim 6 \times 10^{19}$~eV, the
spectrum of ultra-high energy cosmic rays (UHECR) is suppressed
significantly below the power law form that holds at lower
energies. At such high energies, protons are expected to interact
catastrophically with photons in the cosmic microwave background (CMB)
with an attenuation length of ${\cal O}(100)$ Mpc -- the
Greisen-Zatsepin-Kuzmin (GZK)
`horizon'~\cite{Greisen:1966jv,Zatsepin:1966jv,Stecker:1968uc}.
However the observed attenuation does not necessarily imply that the
UHECR are protons since heavy nuclei too will lose energy through
photodisintegration by photons of the cosmic infrared background (CIB)
with a similar energy loss length \cite{Stecker:1969fw,Puget:1976nz}.

The possibility that UHECRs contain a substantial fraction of heavy or
intermediate mass nuclei is interesting to consider for two reasons.
Firstly, recent measurements of UHECR shower profiles at the
PAO~\cite{Unger:2007mc} reveal a gradual decrease in the the average
depth of shower maximum above $\sim 2 \times 10^{18}$~eV, indicating
increasing dominance by heavy nuclei at the highest
energies. Secondly, the maximum energy to which cosmic rays can be
accelerated in ther sources is proportional to their electric charge,
making it less challenging for plausible astrophysical accelerators to
produce heavy nuclei at the highest observed
energies~\cite{Hillas:1985is}.

Recently, the PAO collaboration has also found a correlation between
the arrival directions of UHECRs with energies greater than $6 \times
10^{19}$ eV and nearby active galactic nuclei within $\sim
75$~Mpc~\cite{Cronin:2007zz}. This does suggest that most of these
particles are not deflected by galactic or extragalactic magnetic
fields by more than a few degrees.  As cosmic ray nuclei have higher
electric charge than protons and consequently suffer bigger
deflections by magnetic fields, it is tempting to use this observation
to argue in favor of a proton-dominated UHECR spectrum. UHECR nuclei
can however be significantly disintegrated during propagation, leading
to a population of lighter particles observed at Earth, thus reducing
the impact of magnetic fields (especially in the Galaxy) on UHECR
deflection. Furthermore, there are considerable uncertainties in the
magnitude and structure of galactic and extragalactic magnetic fields.
Hence it is still pertinent to consider the possibility that the
primary cosmic rays are heavy or intermediate mass nuclei and study
their propagation through radiation backgrounds, in order to quantify
the expected composition and spectrum at Earth.

After a long hiatus, there has been quite a bit of work in the past
decade on the propagation of UHECR
nuclei~\cite{Anchordoqui:1997rn,Epele:1998ia,Stecker:1998ib,Bertone:2002ks,Yamamoto:2003tn,Hooper:2004jc,Ave:2004uj,Khan:2004nd,Armengaud:2004yt,Allard:2005ha,Sigl:2005md,Allard:2005cx,Hooper:2006tn,Harari:2006uy,Allard:2007gx,Anchordoqui:2007fi,Arisaka:2007iz}
but we propose here a different approach to the problem. While the
complex process of the photodisintegration of UHECR nuclei into
lighter nuclei and nucleons has so far been addressed using Monte
Carlo techniques, we develop an {\em analytic} description of this
phenomenon. Our primary motivation is to reduce the computation time
needed to calculate the observed UHECR spectrum and composition for a
given injected spectrum, composition and cosmic ray source
distribution.

The rest of this article is organized as follows. In
Sec.~\ref{approach}, we outline the coupled differential equations
governing the population of nuclear species during the
photodisintegration process and find an analytic solution through
physically justified simplifying assumptions. In
Sec.~\ref{cascade_results} we apply this to a cascade initiated by
ultra-high energy iron nuclei, obtaining expressions for the spatial
distribution of each of the species populated in the cascade. In
Sec.~\ref{comparison}, we compare the results of our analytic approach
to those obtained using Monte Carlo methods and demonstrate their
close agreement. In Sec.~\ref{conclusion} we present our conclusions.

\section{The analytic formulation}
\label{approach}

As UHECR nuclei propagate through intergalactic space, they interact
with cosmic radiation backgrounds, fragmenting into lighter nuclei and
nucleons at a rate:
\begin{equation}
 R_{A, Z, i_p, i_n} = \frac{A^2 m^2_p c^4}{2E^2} \int^{\infty}_{0} 
 \frac{\mathrm{d} \epsilon\, n (\epsilon)}{\epsilon^2} 
 \int^{2E\epsilon/A m_p c^2}_{0} 
 \mathrm{d} \epsilon^\prime \epsilon^\prime 
 \sigma_{A, Z, i_p, i_n} (\epsilon^\prime),
\end{equation}
where $A$ and $Z$ are the atomic number and charge of the nucleus,
$i_p$ and $i_n$ are the numbers of protons and neutrons broken off
from a nucleus in the interaction, $n (\epsilon)$ is the density of
background photons of energy $\epsilon$, and $\sigma_{A, Z, i_p,
  i_n}(\epsilon^\prime)$ is the appropriate cross section (see
Appendix~\ref{L}).

An exact analytic treatment of the propagation of UHECR nuclei would
need to take into account all of the many possible decay chains ({\it
  i.e.} all values of $i_p$ and $i_n$ for each nuclear species), as we
have done previously using a Monte Carlo technique
\cite{Hooper:2004jc,Hooper:2006tn,Anchordoqui:2007fi}. The resulting
differential equations are non-trivial to solve, and there would be no
significant computational advantage over the Monte Carlo approach.
Denoting the number of nuclei with atomic number $A$ by $N_A$, the
differential equation describing the population of this state is given
by:
\begin{equation}
 \frac{\mathrm{d}N_A}{\mathrm{d}L} + \frac{N_A}{l_{A \rightarrow A-1}} 
 + \frac{N_A}{l_{A \rightarrow A-2}} + \ldots = 
 \frac{N_{A+1}}{l_{A+1 \rightarrow A}} + 
 \frac{N_{A+2}}{l_{A+2 \rightarrow A}}+ \ldots \ ,
\label{diff.eqn}
\end{equation}
where $L$ is the distance traveled and $l_{i\rightarrow j}$ is the
interaction length for the state $i$ to disintegrate into state $j$
(For a description of how the interaction lengths, $l_{i \rightarrow
  j}$, are calculated, see Appendix~\ref{L}). To begin with, we
consider the simplified case in which only single nucleon loss
processes occur.  This reduces the number of states in the system
dramatically, along with the number of possible transitions.
Eq.~(\ref{diff.eqn}) then simplifies to:
\begin{equation}
\frac{\mathrm{d}N_A}{\mathrm{d}L} + \frac{N_A}{l_A} = 
\frac{N_{A+1}}{l_{A+1}}\ ,
\end{equation}
where a slight change of notation has been introduced --- what was
written earlier as $l_{A \rightarrow A-1}$ is now denoted simply as
$l_A$ (since a particle in state $A+1$ may decay only into state $A$
in our simple model).

The solution to this set of coupled differential equations,
constrained by the initial conditions $N_{n}(L=0)\neq 0$ and
$N_{A}(L=0) = 0$ (for $A \neq n$), is given by (see
Appendix~\ref{solve}):
\begin{equation}
 \frac{N_{A}(L)}{N_{n}(0)} =
 \sum_{m=A}^{n}l_A l_m^{n-A-1} 
 \exp\left(-{\frac{L}{l_m}}\right)\prod_{p=A(\neq m)}^n \frac{1}{l_m -l_p}\ ,
\label{relation} 
\end{equation}
where the interaction lengths, $l_A$, denote those for single
nucleon loss ($A \rightarrow A-1$).

The degree of agreement of this result with that obtained by Monte
Carlo can be improved greatly by redefining $l_A$ in terms of an 
effective interaction length:
\begin{equation}
 L_A = \left(\sum_{n = 1}^{A - 1} 
 \frac{n}{l_{A \rightarrow A-n}}\right)^{-1},
\end{equation}
which accounts for multi-nucleon loss processes as well. 

From Eq.~(\ref{relation}), the average number of various nuclear
species can be obtained as a function of the distance travelled by the
parent cosmic ray. At the high energies of interest, we can sensibly
ignore energy losses due to pair creation (negligible relative to
photodisintegration) and the cosmological redshift (negligible for the
nearby sources which dominate). Hence we can simply relate the energy
of the nuclei to the energy $E_n$ of their parent particle (of mass
number $n$): $E_A = E_n (A/n)$. From now on we will denote the initial
particle energy simply as $E$ so the fraction of particles in state
$A$ with energy $E_A (= EA/n)$ after the parent particle (of mass
number $n$) has propagated a distance $L$, is given by
\begin{equation}
\frac{N_A (L, E_A)}{N_n (0, E)} = 
\sum_{m=A}^{n} L_{A} (E_A) L_m (E_m)^{n-A-1} 
\exp\left(-{\frac{L}{L_m(E_m)}}\right)
\prod_{p = A (\neq m)}^n \frac{1}{L_m (E_m) - L_p (E_p)} 
\label{frac_q}.
\end{equation}
An analogous equation has been used to describe near-threshold pion
production by ultra-high energy protons propagating through the
cosmic microwave background (CMB)~\cite{Aharonian:1990}.

\section{The number and spatial distribution of secondary nuclei and protons}
\label{cascade_results}

In each photodisintegration interaction, a proton is produced
(neutrons are produced too but decay quickly into protons). The
differential equation governing the population of these protons,
denoted by $N_1$, is:
\begin{equation}
\frac{\mathrm{d}N_{1}(E)}{\mathrm{d}L} = 
\frac{N_n (L, E_n)}{L_n (E_n)} + \frac{N_{n-1} (L, E_{n-1})}{L_{n-1}(E_{n-1})} 
 + \ldots \frac{N_2(L, E_2)}{L_2(E_2)}\ ,
\label{protons}
\end{equation}
where $E = E_n/n$. Note that all protons produced during a cascade
initiated by a primary cosmic ray of mass number $n$ with energy $E$ have an
energy $E/A_n$. Their energy distribution is however subsequently altered through interactions with the cosmic background photons.

Integrating Eq.~(\ref{protons}) over $L$, we obtain an expression for $N_1$,
at distance $L'$ and energy $E_1 (= E/A_n)$:
\begin{equation}
  N_{1}(L', E_1) = \int_{0}^{L'} \mathrm{d}L \sum_{m=2}^{n} 
 \frac{N_m (L, E_m)}{L_m (E_m)}\ .
\end{equation}
From this expression it is clear that over sufficiently large
distances, each of the different species contributes equally to the
proton injection spectrum, providing $n$ protons in total. Following
from Eq.~(\ref{relation}), the distribution of each species at
a given distance and energy is given by:
\begin{equation}
\frac{N_q (L, E_q)}{N_n (0, E)} =
\sum_{m=q}^{n} L_q (E_q) L_m (E_m)^{n-q-1} 
\exp\left({\frac{-L}{L_m (E_m)}}\right) 
\prod_{p = q (\neq m)}^{n}
\frac{1}{L_m (E_m) - L_p (E_p)}\ ,
\label{relation2}
\end{equation}
where $E_m = E(A_m/A_n)$ and $E_p = E(A_p/A_n)$.

\begin{figure}[t]
\centering\leavevmode
\includegraphics[width=2.2in,angle=-90]{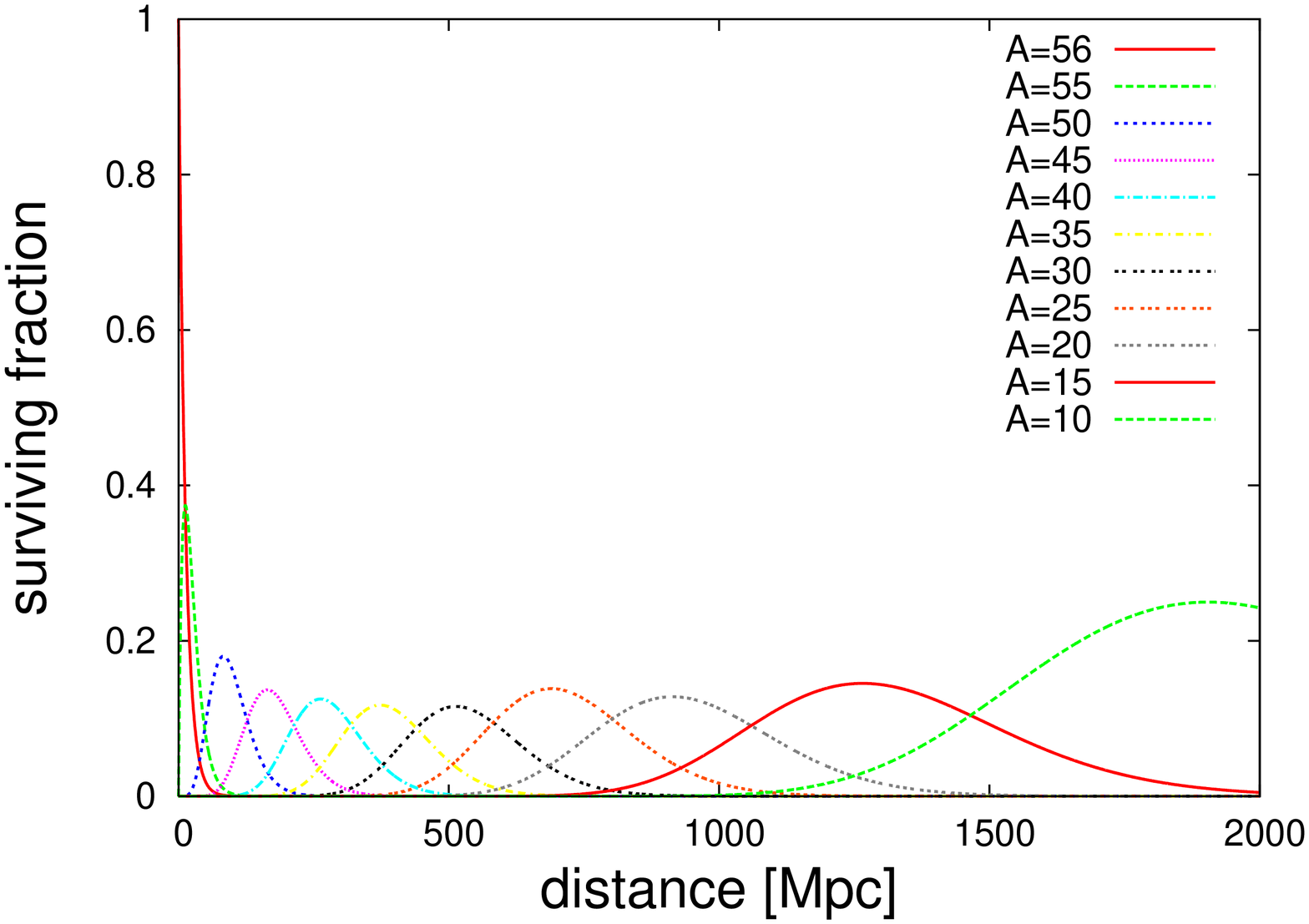}
\includegraphics[width=2.2in,angle=-90]{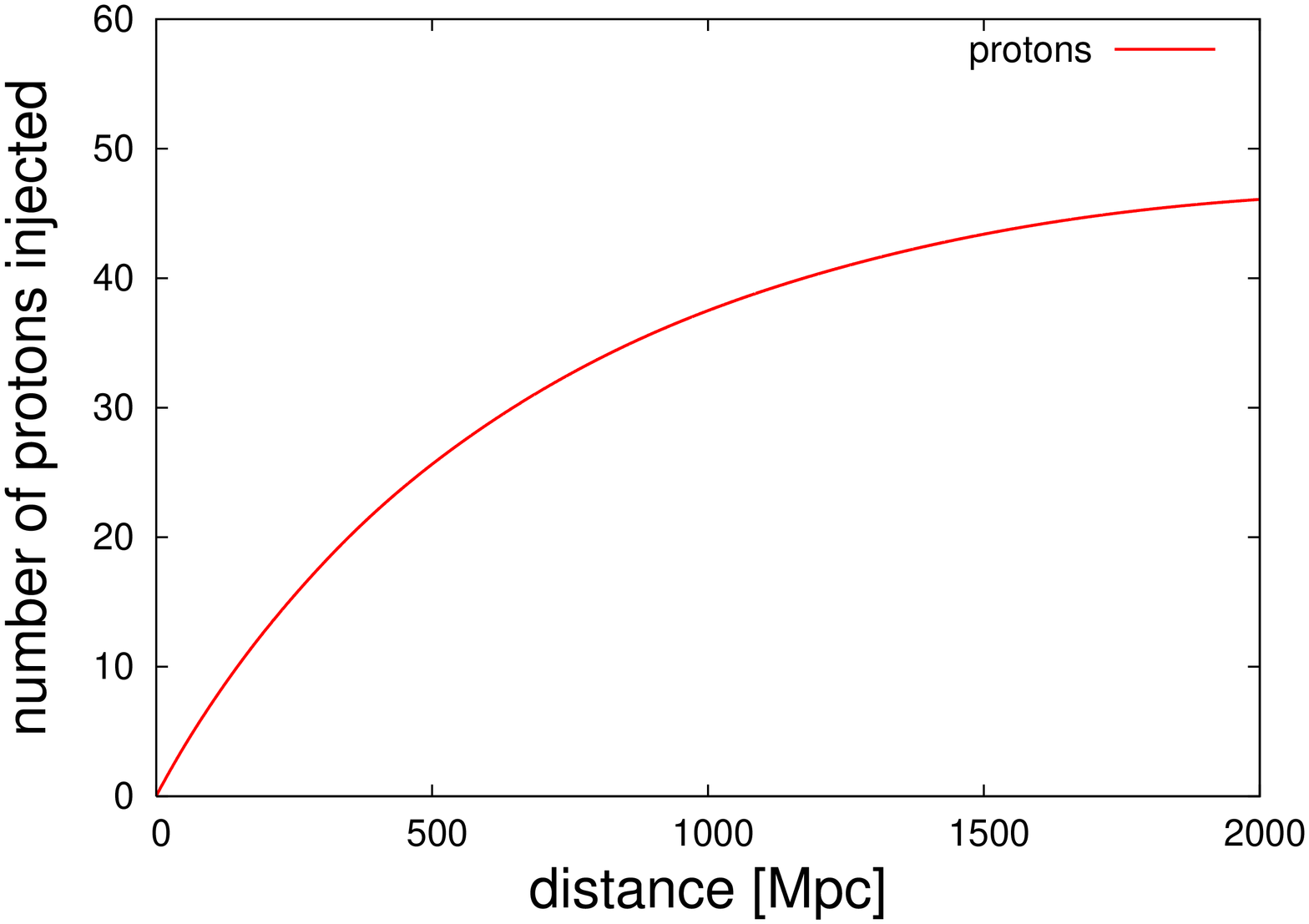}
\caption{The average surviving fraction of various nuclei (left panel)
  and the average number of nucleons disassociated from each primary
  nucleus (right panel) as a function of the distance propagated from
  an extragalactic source of $10^{20}$ eV iron nuclei.}
\label{Fractions}
\end{figure}

As an illustration we apply this to the specific case of 10$^{20}$ eV
iron nuclei, adopting the same CIB spectrum as in our previous work
\cite{Hooper:2004jc}. The surviving fraction of various nuclear
species as a function of distance from the source is shown in the
left frame of Fig.~\ref{Fractions}, while the right frame shows the
average number of nucleons broken off the injected nuclei. From such
distribution functions, the average composition of cosmic rays
arriving at Earth from any set of sources with a specified spatial
distribution and energy spectrum can easily be derived.

Focussing now on the distribution of all particles with a given
energy, $E$, the distribution of such particles is just the sum of the
terms in Eq.~(\ref{relation2}) over all species (requiring different
energy parent nuclei at source):
\begin{equation}
 \frac{\mathrm{d}N_\mathrm{all}(L,E)}{\mathrm{d}L} = 
 \sum_{q=l}^{n} \frac{N_q (L, E_q)}{N_n (0, E_n)}\ ,
\end{equation}
where now $E_q = E(A_n/A_q)$ (note the inversion).

If the particles are injected over a range of distances $L$, from 0
to $L_\mathrm{max}$, with a distribution described by the normalized
function $d (L)$, then the number of a particular species, $q$, at an
energy, $E$, is given by:
\begin{equation}
 N_q (E_q) = \int_{0}^{L_\mathrm{max}} \mathrm{d}L
 \frac{N_q (L, E_q)}{N_n (0, E_n)} d (L).
\label{species_number}
\end{equation}

If particles are injected with a differential energy spectrum
$\mathrm{d}N/\mathrm{d}E \propto E^{-\alpha}$, the total number of all
particles with energy $E_q$ is just
\begin{equation}
 N_\mathrm{tot} (E_q) = \sum_{q=l}^{n} \frac{\mathrm{d}N}{\mathrm{d}E} 
 N_q (E_q) = \sum_{q=l}^{n} \left(\frac{A_n}{A_q}\right)^{1 - \alpha} N_q (E), 
  \label{total_no}
\end{equation}
where $l$ is the lightest species considered in the cascade.

\section{Comparison of the analytic solution with Monte Carlo results}
\label{comparison}

With a knowledge of the effective interaction lengths of each state,
$L_A (E)$, the population of any species of interest after propagating
over a specified distance can be determined following
Eq.~(\ref{frac_q}).  We will now compare the results of
this approach with those from the Monte Carlo calculation
described in our previous work~\cite{Hooper:2004jc}.

\begin{figure}[t]
\centering\leavevmode \mbox{
\includegraphics[width=2.0in,angle=-90]{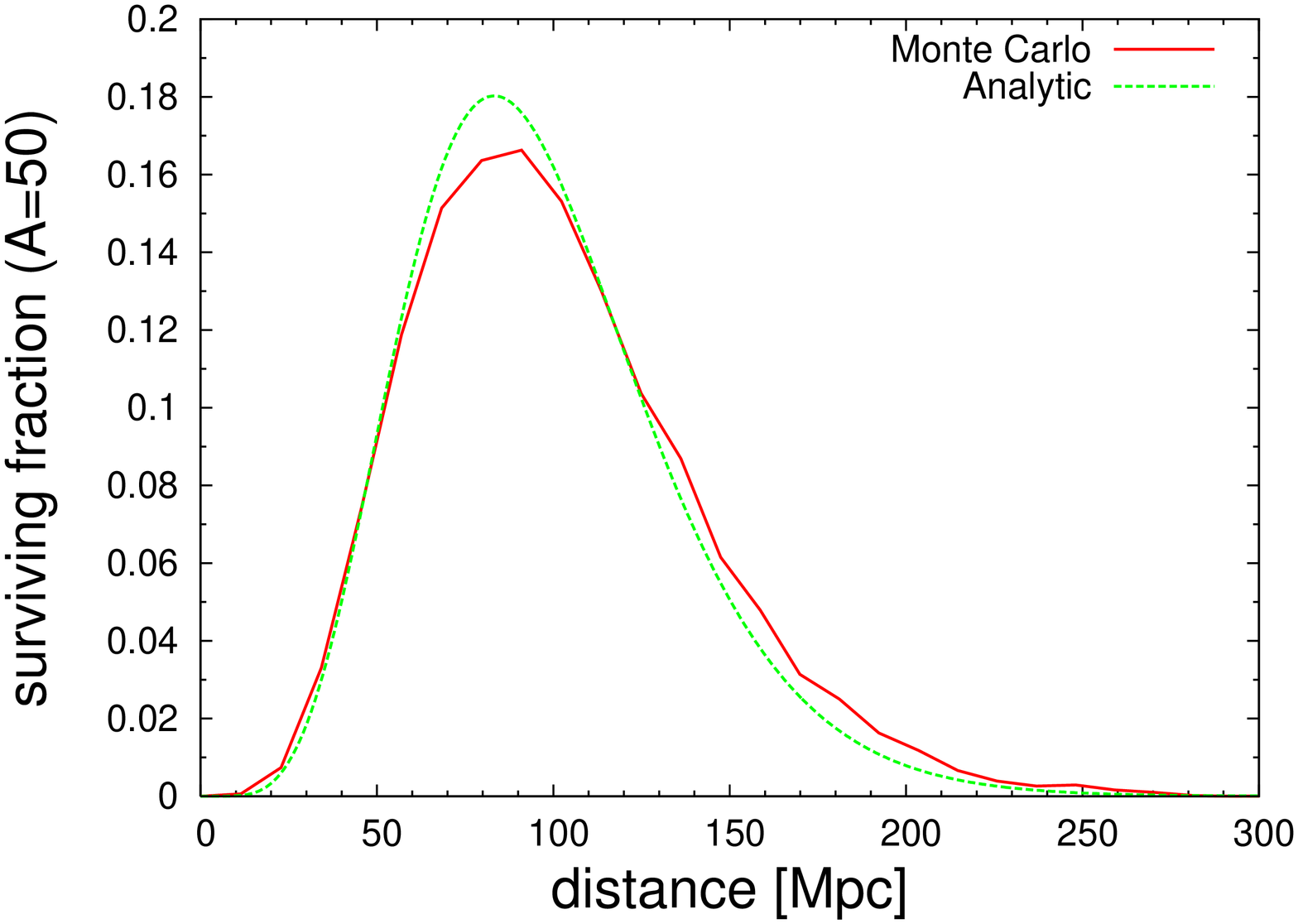}
\includegraphics[width=2.0in,angle=-90]{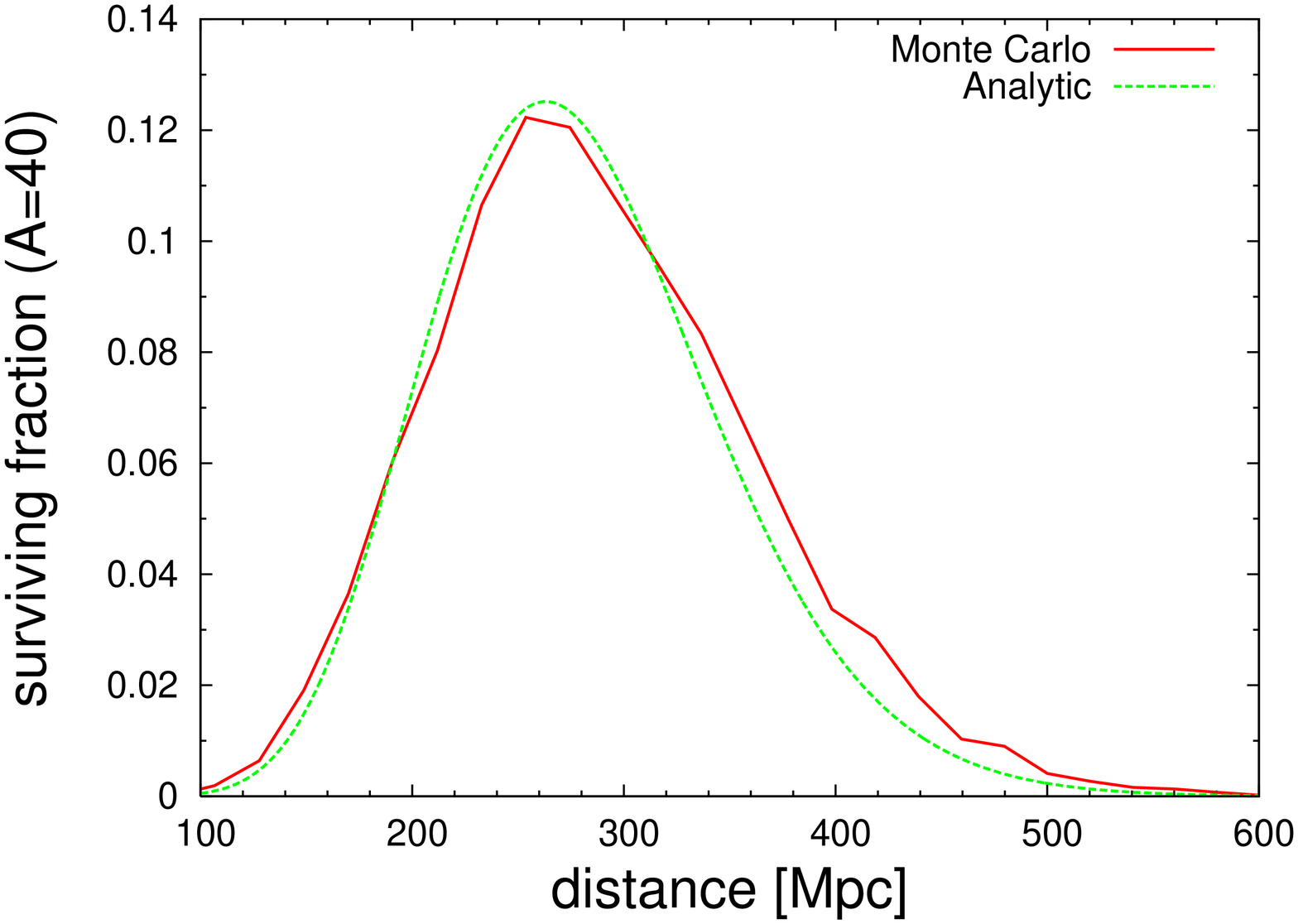} }
\mbox{
\includegraphics[width=2.0in,angle=-90]{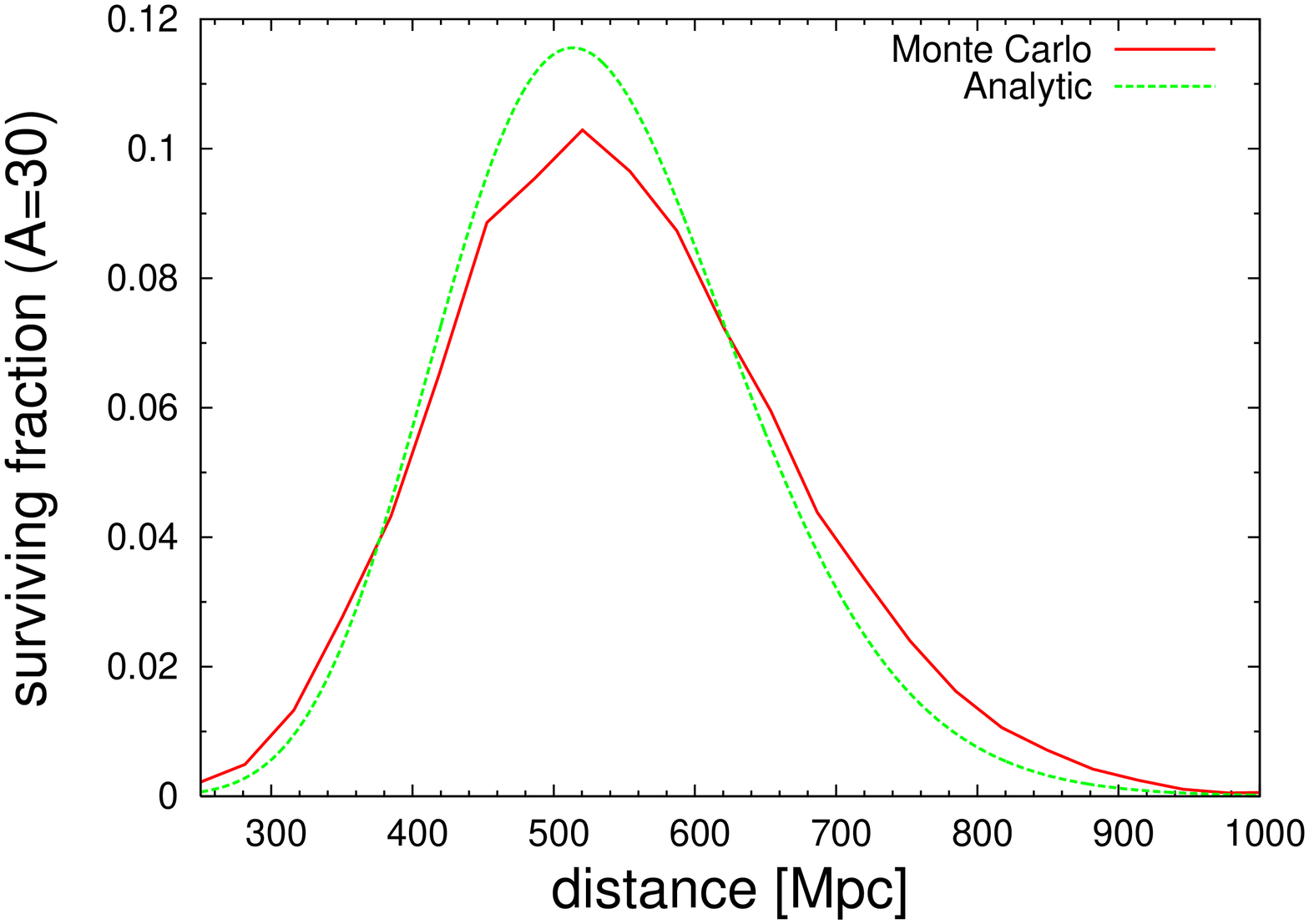}
\includegraphics[width=2.0in,angle=-90]{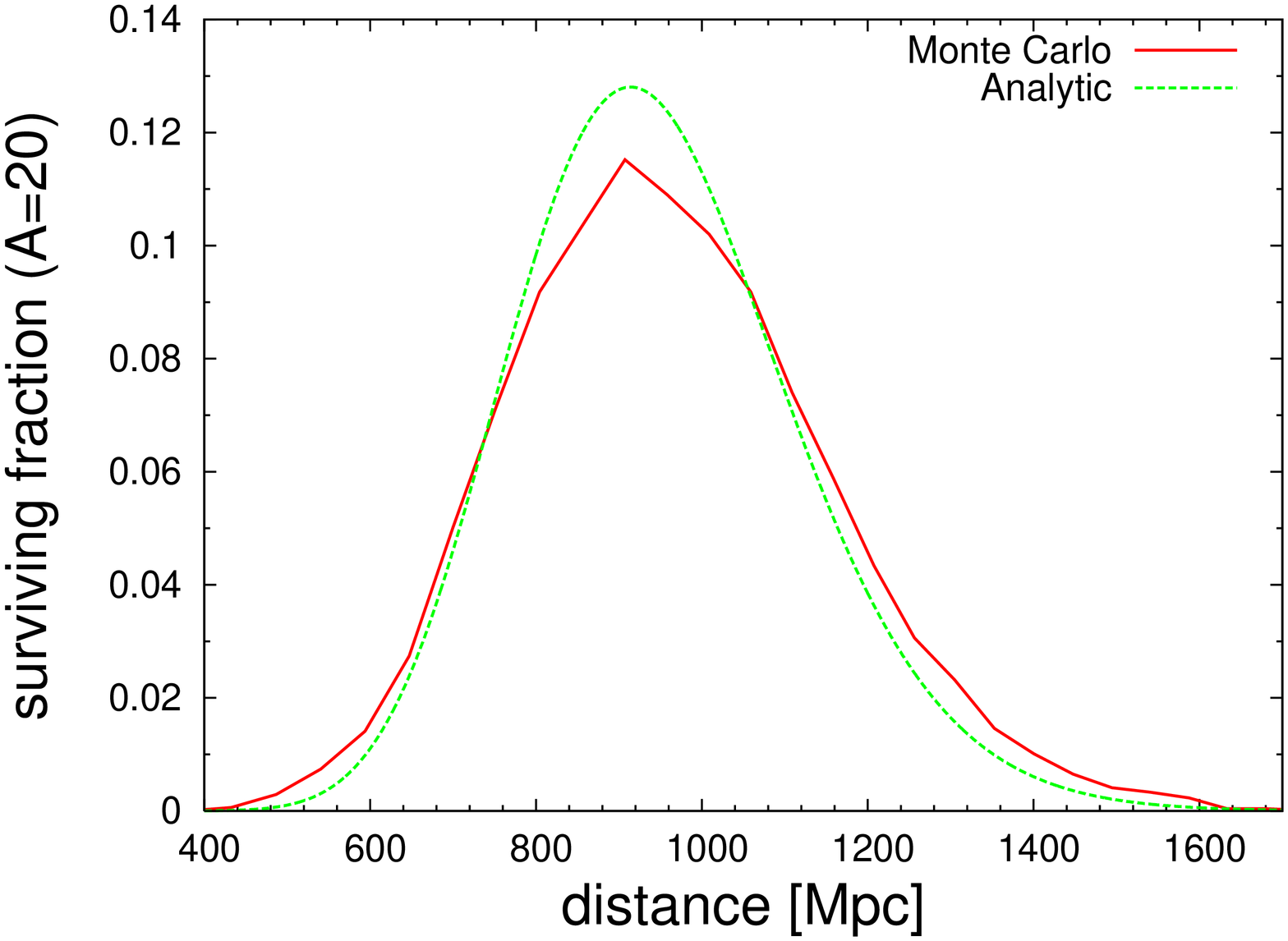} }
\caption{The fraction of particles which start out as $10^{20}$ eV
  iron nuclei in representative states (A=50, 40, 30, and 20) as a
  function of the distance propagated, calculated using both analytic
  and Monte Carlo techniques.}
\label{Comparison}
\end{figure}

In Fig.~\ref{Comparison}, we show the population of five
representative nuclear species, as a function of distance from the
source, as calculated using both analytic and Monte Carlo techniques.
We find excellent agreement, especially for heavier nuclear species.
The small degree of disagreement for lighter nuclei seen in
Fig.~\ref{Comparison} is the result of multi-nucleon loss processes,
the effect of which is most significant over long decay chains.

\begin{figure}[t]
\centering\leavevmode
\includegraphics[width=2.2in,angle=-90]{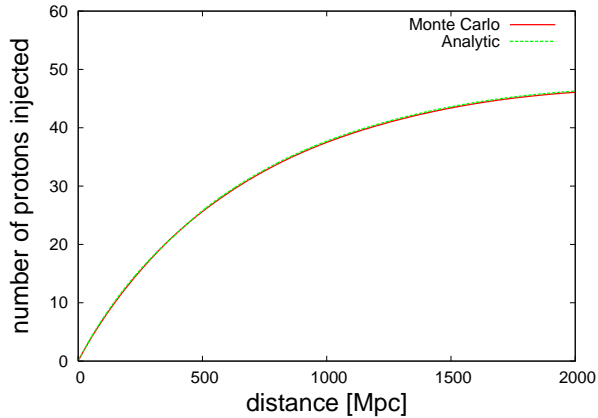}
\caption{The number of protons produced through the
  photodisintegration of $10^{20}$ eV iron nuclei, as a function of
  the distance propagated, calculated using both analytic and Monte
  Carlo techniques.}
\label{Comparison2}
\end{figure}

We continue the comparison of the analytic and Monte Carlo methods in
Fig.~\ref{Comparison2}, where we plot the number of protons produced
through photodisintegration as a function of the distance propagated.
The results from the two methods are virtually indistinguishable.

We can also use our analytic method to calculate the average
composition of the UHECR spectrum after propagation. The average mass
number, $\langle{A}\rangle$, of the particles arriving with energy,
$E$, is given by,
\begin{equation}
 \langle{A(E_q)}\rangle = \frac{\sum_{q=1}^n 
 A_q N_q (E_q)}{N_\mathrm{tot}(E_q)}.
\end{equation}
Integrating the spatial distribution function over a given set of
cosmic ray sources, the total number of each species in the cosmic ray
spectrum can be obtained. A calculation of the number of proton
secondaries from photodisintegration processes which arrive at Earth
is slightly more involved, and requires consideration of both the
production of secondary protons through photodisintegration and their
subsequent pair creation and pion production energy losses on cosmic
photon backgrounds. We can estimate the required quantity for protons:
\begin{equation}
 N_{1}(E_1) = \int_{0}^{L_p (E_1)}\mathrm{d}L' N_1 (L',  E_1),
\end{equation}
where $L_p$ is the energy loss length for protons.

\begin{figure}[t]
\centering\leavevmode
\includegraphics[width=2.2in,angle=-90]{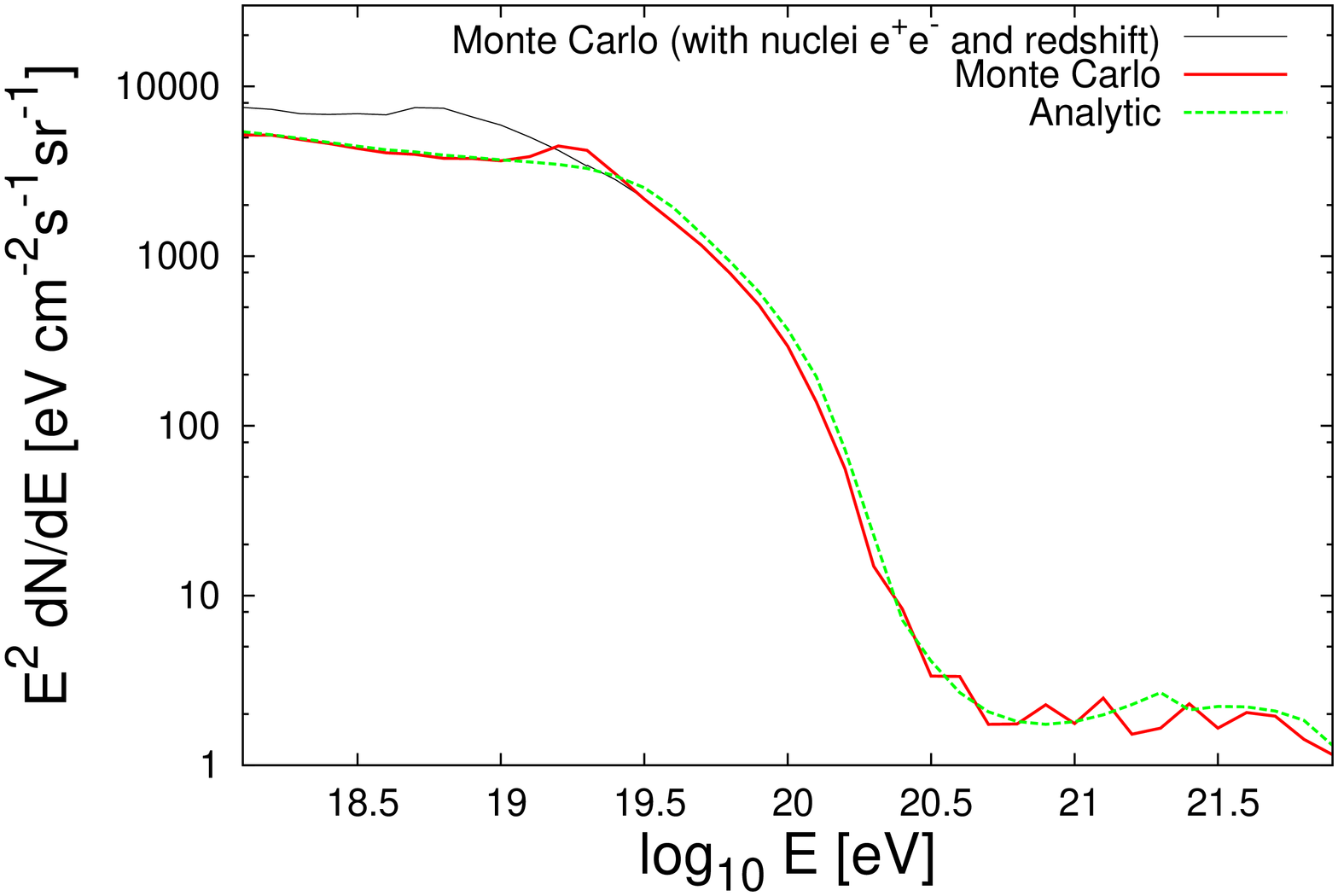}
\includegraphics[width=2.2in,angle=-90]{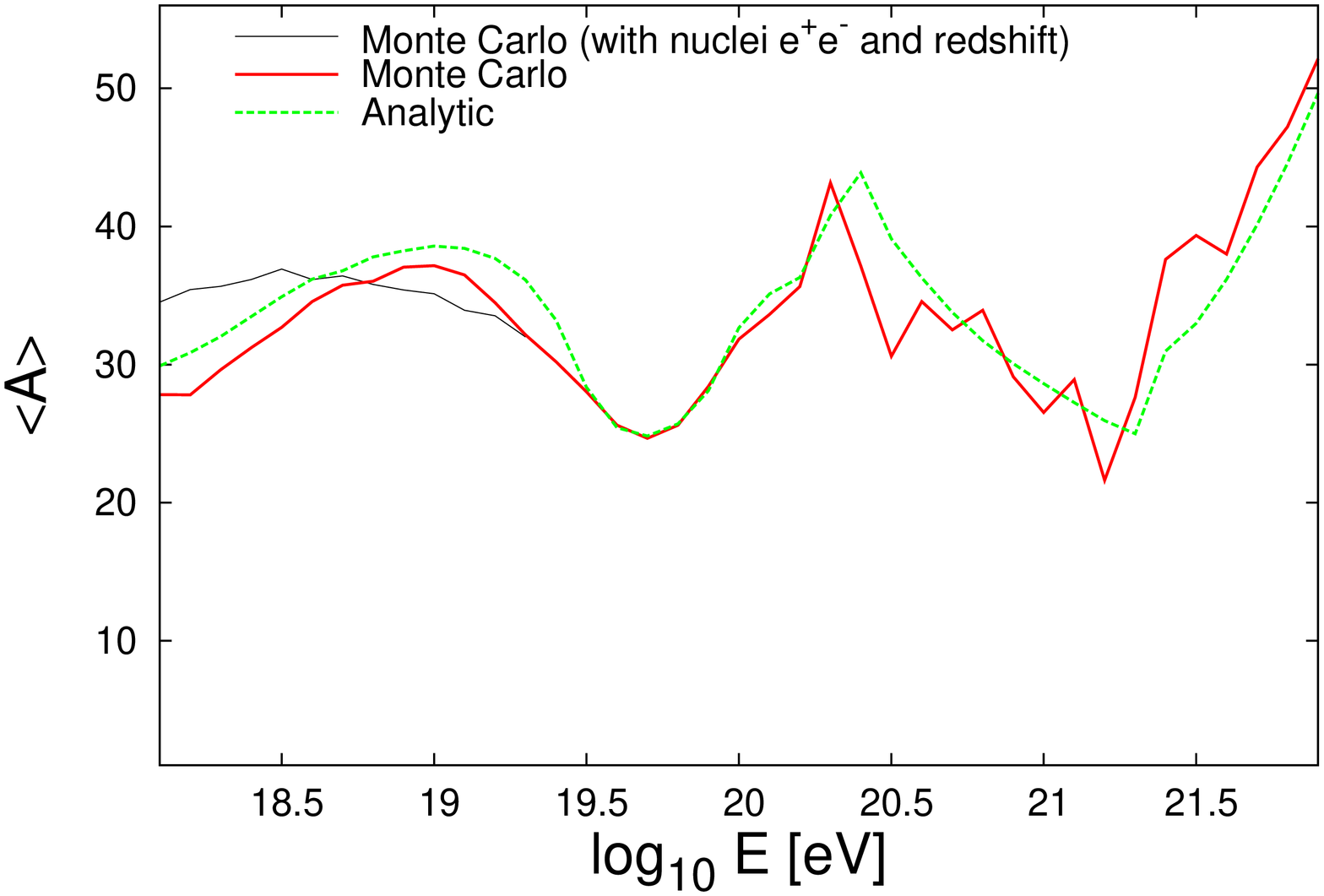}
\caption{The ultra-high energy cosmic ray spectrum (left) and average
  composition (right) calculated using both analytic and Monte Carlo
  techniques, for the case of iron nuclei injected by homogeneously
  distributed sources with $\mathrm{d}N/\mathrm{d}E \propto E^{-2}$ up
  to a maximum energy of $10^{22}$ eV. The black curves show the MC
  results with pair creation losses and redshift effects included.}
\label{Analytic_compare}
\end{figure}

In Fig.~\ref{Analytic_compare}, we compare the spectrum and
composition after propagation found using both our analytic technique
and the Monte Carlo programme, for the case of iron nuclei injected by
a homogeneous distribution of sources with a spectrum
$\mathrm{d}N/\mathrm{d}E \propto E^{-2}$ up to a maximum energy of
$10^{22}$ eV. Due to the energy range of interest in these plots, only
sources up to a redshift $z$=1 were considered. The agreement between
the two techniques is rather good --- the analytic method captures all
of the essential features in the results from the Monte Carlo (in
which pair creation losses and redshift effects were also
included).\footnote{The decrease in $\langle A \rangle$ between
  $10^{20.3}$ and $10^{21.2}$~eV can be understood as follows. Since
  the Fe nuclei are injected to a maximum energy of $10^{22}$~eV, the
  protons produced by photo-disintegration have a maximum energy 56
  times smaller, {\it i.e.} $10^{20.25}$~eV. Above this energy, only
  heavier particles may contribute. As shown in Fig.6 of our previous
  paper~\cite{Hooper:2006tn}, at $10^{20.3}$~eV the interaction
  lengths of low mass nuclei have decreased to a minimum already while
  the interaction lengths of high mass nuclei have just started
  decreasing.  Hence as the energy is increased even further, the
  contribution of the heavier nuclei to the arriving flux decreases,
  hence so does $\langle A \rangle$.} It is seen that the simplifying
approximations made in our analytic approach, namely neglecting pair
creation and redshift energy losses, are quite appropriate for
energies above $\sim10^{19}$~eV.

\section{Conclusion}
\label{conclusion}

We have presented an analytic method to compute the UHECR energy
spectrum and chemical composition at Earth resulting from the
propagation through cosmic radiation backgrounds of heavy nuclei
injected by extragalactic sources. This reduces dramatically the
computation time compared to the usually used Monte Carlo programme
and provides insight into the physical reasons for the results
obtained. These results can be confronted with the observational data,
to enable the identity of the primary particles to be established.

\acknowledgements{DH is supported by the Fermi Research Alliance, LLC
  under Contract No.~DE-AC02-07CH11359 with the US Department of
  Energy. SS acknowledges a STFC Senior Fellowship (PPA/C506205/1) and
  the EU network `UniverseNet' (MRTN-CT-2006-035863). AT acknowledges
  a research stipendium.}

\begin{appendix}
\section{Interaction Lengths}
\label{L}

To carry out the calculations described, first the interaction lengths
for the photodisintegration of nuclei need to be calculated. These are
given by:
\begin{equation}
\label{rate}
\frac{1}{l_{A \rightarrow A-i}} = \frac{A^2 m^2_p c^4}{2E^2}
\int^{\infty}_{0} \frac{\mathrm{d}\epsilon\,n(\epsilon)}{\epsilon^2} 
\int^{2E\epsilon/Am_p c^2}_{0} \mathrm{d}\epsilon^\prime\epsilon^\prime 
\sigma_{A, i} (\epsilon^\prime),
\end{equation}
where $A$ is the atomic number of the nucleus, $i$ is the number of
nucleons broken off from a nucleus in the interaction, $n (\epsilon)$
is the differential number density of background photons of energy
$\epsilon$, and $\sigma_{A, i}(\epsilon^\prime)$ is the appropriate
cross section. We model the photodisintegration cross sections with
the parameterization of
Refs.~\cite{Stecker:1969fw,Puget:1976nz,Stecker:1998ib}:
\begin{equation} 
 \sigma_{A, i} (\epsilon) = \left\{\begin{array}{l@{\quad}l}
 \xi_i \Sigma_\mathrm{d} W_i^{-1} 
 \mathrm{e}^{-2(\epsilon - \epsilon_{p, i})^2/\Delta_i^2}
 \Theta_{+} (\epsilon_\mathrm{thr}) \Theta_{-}
 (\epsilon_1), & \epsilon_\mathrm{thr} \le \epsilon 
 \le \epsilon_1,\quad i = 1, 2 \\
 \zeta\Sigma_\mathrm{d}
 \Theta_{+}(\epsilon_{1})\Theta_{-}(\epsilon_{\rm max})/
(\epsilon_{\rm max}-\epsilon_{1}), &  \epsilon_{1}< \epsilon \le
\epsilon_{\rm max} \\
0, & \epsilon> \epsilon_{\rm max} 
\end{array} \right.
\end{equation}
where $\xi_i$, $\zeta$, $\epsilon_{p, i}$ and $\Delta_i$ are
parameters whose values are obtained by fitting to nuclear data and
the integrated cross-section is 
\begin{equation}
\Sigma_\mathrm{d} \equiv \int_{0}^{\infty} \sigma(\epsilon)\,d\epsilon 
= 2\pi^2 \frac{e^2}{4\pi \epsilon_0} \frac{\hbar c}{m_p c^2}\frac{(A - Z)Z}{A} 
= 60 \frac{(A - Z)Z}{A}
\mbox{mb-MeV},
\end{equation}
with $Z$ being the charge of the nucleus. The function $W_i$ is given by:
\begin{equation}
 W_i = \Delta_i \sqrt{\frac{\pi}{8}}
 \left[\mbox{erf}\left(\frac{\epsilon_\mathrm{max} - \epsilon_{p, i}}
 {\Delta_i/\sqrt{2}}\right) + \mbox{erf}\left(\frac{\epsilon_{p, i} 
 - \epsilon_1}{\Delta_{i}/\sqrt{2}}\right)\right].
\end{equation}
Here, $\Theta_{+}(x)$ and $\Theta_{-}(x)$ are the Heaviside step
functions, $\epsilon_1$ = 30 MeV, $\epsilon_{\rm max}$ = 150 MeV, and
the threshold energy for a given process is in most cases is
$\epsilon_{\rm thr} \approx i \times 10$ MeV (values are tabulated in
Ref.~\cite{Stecker:1998ib}). These cross sections are dominated by the
giant dipole resonance which peaks in the energy range $\sim 10-30$
MeV; at higher energies, quasi-deuteron emission is the main process.

%

Nuclei with $A=5-9$ are very unstable and quickly decay to helium and
lighter elements. We approximate this behaviour by setting $l_{A
  \rightarrow 4}=0$ for $A=5-9$.

\section{Development of a Solution}
\label{solve}

\noindent
Consider the differential equation,
\begin{equation}
 \frac{\mathrm{d}N_A}{\mathrm{d}l} + \frac{N_A}{l_A} = 
 \frac{N_{A + 1}}{l_{A + 1}}.
\end{equation}
This may be written as,
\begin{equation}
 \exp\left(-{\frac{l}{l_A}}\right)\frac{\mathrm{d}}{\mathrm{d}l} 
 \exp\left({\frac{l}{l_A}}N_A\right) = \frac{N_{A + 1}}{l_{A + 1}},
\end{equation}
so that
\begin{equation}
 N_l = \exp\left(-{\frac{l}{l_A}}\right) \int \mathrm{d}l
 \exp\left({\frac{l}{l_A}}\right)\frac{N_{A+1}}{l_{A+1}}
\label{N_A},
\end{equation}
with the initial conditions of the system being, $N_n (L = 0) \neq 0$
and $N_A (L = 0) = 0$ (for $A \neq n$).

Let us assume that the solution is:
\begin{equation}
 \frac{N_A(l)}{N_n(0)} = \sum_{m=A}^{n} l_A l_m^{n-A-1} 
 \exp\left(-{\frac{l}{l_m}}\right)\prod_{p=A(\neq m)}^n \frac{1}{l_m - l_p},
\label{Assume_Solution}
\end{equation}
where $n$ is the initial state the particles are in.  The following
will be a `proof by induction'.

If the above is true for $A$, then it is true also for $A+1$:
\begin{equation}
 \frac{N_{A+1}(l)}{N_n(0)} = \sum_{m=A+1}^{n} L_{A+1}l_m^{n-A-2}
 \exp\left(-{\frac{l}{l_m}}\right) \prod_{p = A + 1 (\neq m)}^n
 \frac{1}{l_m - l_p},
\end{equation}
which using Eq.~(\ref{N_A}) gives
\begin{eqnarray}
 \frac{N_A (l)}{N_n (0)} &=& \exp\left(-{\frac{l}{l_A}}\right)
 \int \mathrm{d}l \exp\left({\frac{l}{l_A}}\right) 
 \sum_{m = A + 1}^{n} l_m^{n - A - 2} 
 \exp\left(-{\frac{l}{l_m}}\right) 
 \prod_{p = A +1 (\neq m)}^n \frac{1}{l_m - l_p} \nonumber\\
 &=& \sum_{m = A + 1}^n l_m^{n - A - 2} 
 \left[\left(\frac{1}{l_A} - \frac{1}{l_m}\right)^{-1} 
 \exp\left(-{\frac{l}{l_m}}\right)\right]\prod_{p = A + 1 (\neq m)}^n 
 \frac{1}{l_m - l_p} - c\exp\left(-{\frac{l}{l_A}}\right) \nonumber\\
 &=& \sum_{m = A + 1}^n l_A l_m^{n - A - 1} 
 \exp\left(-{\frac{l}{l_m}}\right) 
 \prod_{p = A (\neq m)}^n \frac{1}{l_m - l_p} - 
 c\exp\left(-{\frac{l}{l_A}}\right), 
\label{new_expression}
\end{eqnarray}
Applying the boundary condition $N_{A}(L=0)=0$, we get
\begin{equation}
 c = \sum_{m = A + 1}^n l_A l_m^{n-A-1} 
 \prod_{p = A (\neq m)}^n \frac{1}{l_m - l_p}.
\end{equation}
For Eq.~(\ref{Assume_Solution}) and Eq.~(\ref{new_expression}) to be
equivalent, it is necessary that,
\begin{equation}
 \sum_{m = A}^n l_A l_m^{n-A-1} 
\prod_{p = A (\neq m)}^n \frac{1}{l_m - l_p} = 0 
\label{requirement}
\end{equation}
This final expression may be recognised as an expression of the
`Vandermond determinant' with the final column repeated. Since this is
indeed true, our assumed solution \ref{Assume_Solution} must also be
correct.

\end{appendix}

\end{document}